\documentclass[final,conference, twoside, 10pt, letterpaper]{IEEEtran}
\usepackage{Bamble}
\usepackage{Bmacros}
\usepackage[noadjust]{cite}
\usepackage{accents}

\newcommand{\cev}[1]{\overset{\leftarrow}{#1}}
\newcommand{\longcev}[1]{\overleftarrow{#1}}
\newcommand{\ours}{\mathtt{s}}
\newcommand{\ourp}{\mathtt{p}}
\newcommand{\ourS}{\mathtt{S}}
\newcommand{\ourP}{\mathtt{P}}
\newcommand{\inferior}{\sqsubseteq}
\newcommand{\improved}{\sqsupseteq}
\newcommand{\degraded}{\overset{\text{d}}{\sqsubseteq}}
\newcommand{\permutation}{\overset{\text{p}}{\sqsubseteq}}
\newcommand{\joker}[1]{\overset{\text{r}_{#1}}{\sqsubseteq}}
\newcommand{\jokerA}[1]{\overset{\text{r}^{A}_{#1}}{\sqsubseteq}}
\newcommand{\jokerB}[1]{\overset{\text{r}^{B}_{#1}}{\sqsubseteq}}
\newcommand{\lngs}[1]{\bar{#1}}
\newcommand{\lngp}[1]{\tilde{#1}}
\newcommand{\extend}[1]{\underaccent{\bullet}{#1}}
\newcommand{\chop}[1]{\underaccent{\circ}{#1}}

\newcommand{\twobibs}[2]{#2} %

\begin{document}
\title{Strong Polarization for Shortened and Punctured Polar Codes}
\date{}
\author{\IEEEauthorblockN{Boaz~Shuval, Ido~Tal\\
The Andrew and Erna Viterbi Faculty of Electrical and Computer Engineering,\\
Technion, Haifa 32000, Israel.\\
Email: \{\texttt{bshuval@}, \texttt{idotal@ee.}\}\texttt{technion.ac.il}}}

\maketitle

\begin{abstract}
	Polar codes were originally specified for codelengths that are powers of two. In many applications, it is desired to have a code that is not restricted to such lengths. Two common strategies of modifying the length of a code are shortening and puncturing. Simple and explicit schemes for shortening and puncturing were introduced by Wang and Liu, and by Niu, Chen, and Lin, respectively. In this paper, we prove that both schemes yield polar codes that are capacity achieving.
	Moreover, the probability of error for both the shortened and the punctured polar codes decreases to zero at the same exponential rate as seminal polar codes. These claims hold for \emph{all} codelengths large enough.
\end{abstract}
\section{Introduction}
Polar codes \cite{Arikan:09p} are based on a recursive transform, yielding codes whose codelengths are powers of two. They have been proven to achieve the capacity of many channel settings \cite{HondaYamamoto:12p,Sasoglu:12c,SasogluTal:19p,ShuvalTal:19.2p,TPFV:22p,PfisterTal:21c,Hof+:13p,HofShamai:10a,ARTKS:10p,MahdavifarVardy:11p,Mondelli+:15p,Mondelli+:18p,TFV:21p,SasogluWang:16p,ShuvalTal:18a,AbbeTelatar:10p,STY:13p,Goela+:15p,Mahdavifar:20p,AravaTal:23c}.
Often, it is desirable to transmit a message whose length is not limited to a power of $2$. Shortening and puncturing~\cite[Problems 2.3 and 2.14]{Roth:06b}, \cite[Chapter 1\S 9]{MacWilliamsSloane:77b} are two common methods of reducing the length of a given code. Such methods were extensively studied for polar codes, see~\cite{NCL:13c,WangLiu:14p,BGL:17c,OliveiraLamare:18p,OliveiraLamare:18c,TCG:19c,YaoMa:23p} and the references therein. In this paper, we focus on the puncturing method of~\cite{NCL:13c} and the shortening method of~\cite{WangLiu:14p}\footnote{The title of \cite{WangLiu:14p} claims a puncturing method, but in fact describes a shortening method.}.  In the sequel, for brevity, we will refer to transforms based on these methods as the ``shortening transform'' and ``puncturing transform,'' respectively. We show that these schemes achieve capacity, with probability of error decreasing at the same exponential rate as seminal polar codes. This holds for all codelengths large enough. For simplicity, we focus on the setting of a binary-input memoryless channel, which may be non-symmetric (BM channel).

The following theorem is a shortened version\footnote{Or is it a punctured version?} of our main result. It will follow as a straightforward corollary of the more general Theorem~\ref{thm:mainAndGeneral}. It assumes a fixed input distribution $p(x)$ and a fixed BM channel $W(y|x)$. We denote by $Z(X|Y)$ and $K(X|Y)$ the conditional Bhattacharyya parameter and the total variation distance, respectively (see \cite[Definitions 2 and 3]{ShuvalTal:19.2p}). Furthermore for $X$ and $Y$ with joint distribution $W(x;y) \triangleq p(x) W(y|x)$, we denote by $H(X|Y)$ the conditional entropy of $X$ given $Y$ and by $H(X)$ the entropy of $X$.
\begin{theorem} \label{thm:main}
	Let $\bv{X}$ be a random vector of length $M$ with i.i.d.\ entries, each sampled from an input distribution $p(x)$. Let $\bv{Y}$ be the result of passing $\bv{X}$ through a BM channel $W(y|x)$. Let $\bv{U}$ of length $M$ be the result of transforming $\bv{X}$ via either the shortening transform or the puncturing transform. Fix $0 < \beta < 1/2$. Then,
\begin{IEEEeqnarray}{rCl}
	\lim_{M \to \infty} \frac{1}{M}\left|\left\{ i : Z(U_i|U^{i-1},\bv{Y}) < 2^{-M^\beta} \right\}\right| &=& 1 - H(X|Y) , \label{eq:mainTheorem:Z} \\
\lim_{M \to \infty} \frac{1}{M}\left|\left\{ i : K(U_i|U^{i-1}) < 2^{-M^\beta} \right\}\right| &=& H(X) . \label{eq:mainTheorem:K}
\end{IEEEeqnarray}
\end{theorem}
The above theorem implies that, similar to the power-of-two setting, we can use successive cancellation and the Honda-Yamamoto scheme \cite{HondaYamamoto:12p} to define a code whose rate approaches $I(X;Y)$ and whose probability of error is upper bounded by $2^{-M^\beta}$ for $0 < \beta < 1/2$ fixed and \emph{all} integer $M$ large enough. Moreover, both encoding and decoding can be calculated in time $O(M \log M)$. 

\section{The shortening and Puncturing Transforms}
In this section we define both the shortening and the puncturing transforms. To do so, for a given codelength $M$, we denote by $N$ the smallest power of two greater or equal to $M$. That is,
\begin{equation}\label{eq:N}
	N = 2^{\lceil \log_2 M \rceil} .
\end{equation}
We also denote
\begin{equation} \label{eq:n}
n = \lceil \log_2 M \rceil= \log_2 N  .
\end{equation}

Since we will make heavy use of bit-reversals, it is natural to use zero-based indexing. That is, an index $0 \leq i < N$ has binary representation $i = \sum_{j = 0}^{n-1} b_j 2^j$. The corresponding vector is $\bv{b} = \begin{bmatrix} b_0 & b_1 & \cdots & b_{n-1} \end{bmatrix}$. The reversed vector is $\cev{\bv{b}}  = \begin{bmatrix} b_{n-1} & b_{n-2} & \cdots & b_{0} \end{bmatrix}$. The corresponding bit-reversed index is $\cev{i} = \sum_{j = 0}^{n-1} b_j 2^{n-1-j}$.

\subsection{Generalization of Key Polar Coding Concepts}
Seminal polar codes revolve around three key concepts:
\begin{itemize}
	\item The polar transform, an invertible transform that transforms a vector $\bv{x}$ of bits to a vector $\bv{u}$ of bits, both of the same length $N = 2^n$.
	\item The `$-$' and `$+$' operations, denoted $\boxast$ and $\circledast$, respectively. They transform two joint distributions $A$ and $B$ into new joint distributions, $A\boxast B$ and $A\circledast B$, respectively. This is a slight generalization of the seminal setting, in which $A$ and $B$ were the same distribution, in which case $A \boxast A$ was denoted $A^-$ and $A\circledast A$ was denoted $A^+$.
	\item The connection between the polar transform and the `$-$' and `$+$' operations.
\end{itemize}
We now briefly review these concepts and show how to generalize them to the shortening and puncturing setting.

\subsubsection{The Polar Transform}\label{subsec:PolarTransform}
The seminal polar transform takes a vector $\bv{x}$ of length $N=2^n$ and produces a transformed vector $\bv{u}$, also of length $N$. A simple way to define this transform is by two operations that take a vector of length $N$ and produce a vector of length $N/2$. Namely,
\begin{multline} \label{eq:x0}
	\begin{bmatrix} x_0 & x_1 & \cdots & x_{N-1} \end{bmatrix}^{[0]} \\
	= \begin{bmatrix} x_0 \oplus x_1 & x_2 \oplus x_3 & \cdots & x_{N-2} \oplus x_{N-1} \end{bmatrix}
\end{multline}
and
\begin{multline} \label{eq:x1}
	\begin{bmatrix} x_0 & x_1 & \cdots & x_{N-1} \end{bmatrix}^{[1]} \\
	= \begin{bmatrix} x_0 \triangleright x_1 & x_2 \triangleright x_3 & \cdots & x_{N-2} \triangleright x_{N-1} \end{bmatrix},
\end{multline}
where\footnote{The notation $\triangleright$ is suggestive of an arrowhead pointing at the output of the operation.}
$\alpha \triangleright \beta = \beta$. We denote for $\bv{b} = \begin{bmatrix} b_0 & b_1 & \cdots & b_{\ell-1} \end{bmatrix}$, 
\begin{equation} \label{eq:xb}
	\bv{x}^{[\bv{b}]} = \left( \cdots \left(\left(\bv{x}^{[b_{0}]}\right)^{[b_{1}]}\right) \cdots \right)^{[b_{\ell-1}]},
\end{equation}
that is, the result of recursively applying $(\cdot )^{[0]}$ and $(\cdot )^{[1]}$ operations. 
Then, entry $i = \sum_{j = 0}^{n-1} b_j 2^j$ of $\bv{u}$ is $\bv{x}^{[\cev{\bv{b}}]}$, 
where $\bv{b} = \begin{bmatrix} b_0 & b_1 & \cdots & b_{n-1} \end{bmatrix}$.

We now extend the definitions of operations $\oplus$ and $\triangleright$ to apply over the set $\{0,1,\ours,\ourp\}$. Here, $\ours$ represents a shortened bit and $\ourp$ a punctured bit. Namely, the generalizations of both operations are given in the following tables, which are to be read as $\alpha \cdot \beta$ with $\alpha$ a row and $\beta$ a column. E.g., $1 \triangleright 0 = 0$. 
\begin{equation}\label{eq:OperationTables}
	\begin{tabular}{ c| c  c  c  c }
$\oplus$ & $0$ & $1$ & $\ours$ & $\ourp$ \\
\hline
$0$ & $0$ & $1$ & $0$ & $\emptyset$ \\ 
$1$ & $1$ & $0$ & $1$ & $\emptyset$ \\ 
$\ours$ & $\emptyset$ & $\emptyset$ & $\ours$ & $\emptyset$ \\ 
$\ourp$ & $\ourp$ & $\ourp$ & $\ourp$ & $\ourp$  
\end{tabular} \quad ;  \quad 
\begin{tabular}{ c| c  c  c  c }
$\triangleright$ & $0$ & $1$ & $\ours$ & $\ourp$ \\
\hline
$0$ & $0$ & $1$ & $\ours$ & $\emptyset$ \\ 
$1$ & $0$ & $1$ & $\ours$ & $\emptyset$ \\ 
$\ours$ & $\emptyset$ & $\emptyset$ & $\ours$ & $\emptyset$ \\ 
$\ourp$ & $0$ & $1$ & $\ours$ & $\ourp$  
\end{tabular}.
\end{equation}
\looseness=-1 
In the above, $\emptyset$ denotes the ``don't care'' value. That is, $\ours$ will never be the first argument, unless the second argument is $\ours$, and $\ourp$ will never be the second argument, unless the first argument is $\ourp$. Also, although this is a setting we do not consider further in this paper, note that the above table implies that we can have both shortened and punctured bits in our codeword.

\subsubsection{The `$-$' and `$+$' Operations}\label{subsec:MinusPlusOperation}
In the seminal setting, the `$-$' and `$+$' operations each transform two identical channels into a new channel. Here, they each transform two joint distributions into a new joint distribution. That is, let $A(x_0;y_0)$ be the joint distribution on the pair $(x_0,y_0) \in \mathcal{X} \times \mathcal{Y}_0$, where henceforth $\mathcal{X} = \{0,1\}$. Further let $B(x_1;y_1)$ be the joint distribution on the pair $(x_1,y_1) \in \mathcal{X} \times \mathcal{Y}_1$. Then,
\begin{IEEEeqnarray}{rCl}
	(A \boxast B)(u_0; y_0, y_1 ) &=& \sum_{x_1 \in \mathcal{X}} A(u_0 \oplus x_1 ;y_0) B(x_1;y_1), \IEEEeqnarraynumspace\label{eq:AboxastBdef} \\[0.1cm]
	(A \circledast B)(u_1; u_0,y_0, y_1 ) &=& A(u_0 \oplus u_1 ;y_0) B(u_1;y_1) . \label{eq:AcircledastBdef} 
\end{IEEEeqnarray}

We now define two special joint distributions, $\ourS$ and $\ourP$, corresponding to a ``shortened'' distribution and a ``punctured'' distribution, respectively. Both $\ourS$ and $\ourP$ are over $\mathcal{X} \times \{?\}$. They are given by 
\begin{IEEEeqnarray}{rCl}
	\ourS(x;y) &=&
	\begin{cases}
		1 , & x = 0, y = ? , \\
		0, & \text{otherwise},
	\end{cases}  \label{eq:ourSDef} \\[0.1cm]
	\ourP(x;y) &=&
	\begin{cases} 
		\frac{1}{2} , & x \in \mathcal{X}, y = ? . %
	\end{cases} \label{eq:ourPDef}
\end{IEEEeqnarray}
For reasons that will become clearer later, we call $\ourS$ the `superb' distribution and $\ourP$ the `pitiful' distribution.

\subsubsection{The Connection between the Polar Transform and the `$-$' and `$+$' Operations}\label{subsec:Connections}
Consider the vector of joint distributions $\bv{A} = \begin{bmatrix} A_0 &  A_1 & \cdots & A_{N-1} \end{bmatrix}$. We define $\bv{A}^{[0]}$, $\bv{A}^{[1]}$, and $\bv{A}^{[\bv{b}]}$  by adapting \eqref{eq:x0}, \eqref{eq:x1}, and \eqref{eq:xb}, respectively. We adapt these by replacing $x_i$ with $A_i$, $\oplus$ with $\boxast$, and $\triangleright$ with $\circledast$.

Let $0 \leq i < N$ with binary representation $i = \sum_{j=0}^{n-1}b_j2^{n-1-j}$. 
Then, there exists an invertible function $f$ such that
\[
	\Prob{U_i = u_i ; U_0^{i-1} = u_0^{i-1}, \bv{Y}=\bv{y}} = \bv{A}^{[\bv{b}]}(u_i; f(u_0^{i-1}, \bv{y})).
\]

\subsection{The Shortening Transform}
For a general (not necessarily polar) code $\mathcal{C}$ of length $N$, shortening is defined through an index set $\mathcal{S}$. Namely, to shorten $\mathcal{C}$, we first consider the subset of codewords $\bv{c} \in  \mathcal{C}$ for which $c_i = 0$ for all $i \in \mathcal{S}$. For every such codeword, since we know the values at the indices $\mathcal{S}$, there is no point in transmitting them. Hence, the shortened code is the above subset, after removing the indices $\mathcal{S}$. Note that the shortened code has length $N - |\mathcal{S}|$.

In the Wang-Liu shortening scheme \cite{WangLiu:14p},
\begin{equation} \label{eq:Sset}
	\mathcal{S} = \{\longcev{N{-}1},\longcev{N{-}2},\ldots,\longcev{N{-}M}\} .
\end{equation}
That is, the \emph{last} $N-M$ bits of the codeword, before bit reversal, are constrained to be $0$. This implies that the last $N-M$ entries of the corresponding transformed vector are frozen to $0$. Successive-cancellation (SC) decoding is performed exactly as for seminal polar codes, save for setting a log-likelihood ratio (LLR) value of infinity to the shortened bits. See~\cite{WangLiu:14p} for details. 

We define the shortening transform of a vector $\bv{x}$ of $M$ bits in two equivalent ways. In the first way, we define a vector $\lngs{\bv{x}}$ of length $N = 2^{\lceil \log_2 M \rceil}$ with indices $\mathcal{S}$ set to $\ours$. We then copy $\bv{x}$ into $\lngs{\bv{x}}$ in order. That is, removing from $\lngs{\bv{x}}$ the indices in $\mathcal{S}$ recovers $\bv{x}$. Next, we compute $\lngs{\bv{u}}$, as explained in \Cref{subsec:PolarTransform}. We note that by the special choice of $\mathcal{S}$, we will never encounter an `$\emptyset$' entry in~\eqref{eq:OperationTables}. Lastly, we define $\bv{u}$ by the result of removing the last $N-M$ entries from $\lngs{\bv{u}}$. We remark in passing that these removed entries were all equal to $\ours$. 

Observe that had we replaced $\ours$ with $0$ in  \eqref{eq:OperationTables}, no contradiction would have arisen. Thus, in the spirit of shortening, had we replaced $\ours$ with $0$ in the extension from $\bv{x}$ to $\lngs{\bv{x}}$, then the last $N-M$ entries in $\lngs{\bv{u}}$ would also have been $0$, and $\bv{u}$ would have been the same as that from the previous paragraph. This is the second way of defining the shortening transform: replace all $\ours$ in the above with $0$.
\begin{remark} \label{rmk:uprefix}
	Note that $\bv{u}$ equals the prefix of length $M$ of $\lngs{\bv{u}}$. That is, for $0 \leq i < M$, $u_i = \lngs{u}_i$.
\end{remark}

\subsection{The Puncturing Transform} \label{subsec:puncturingTransform}
Similar to shortening, for a general code $\mathcal{C}$ of length $N$, puncturing is defined through an index set $\mathcal{P}$. Namely, to puncture $\mathcal{C}$, we simply remove the indices $\mathcal{P}$ from the codeword. The punctured code has length $N - |\mathcal{P}|$.

In the Niu-Chen-Lin puncturing scheme \cite{NCL:13c},
\[
	\mathcal{P} = \{\longcev{0},\longcev{1},\ldots,\longcev{N-M{-}1}\} .
\]
That is, the \emph{first} $N-M$ bits of the codeword, before bit reversal, are removed. This implies that the first $N-M$ entries of the corresponding transformed vector are frozen. Successive-cancellation (SC) decoding is performed exactly as for seminal polar codes, save for setting a log-likelihood ratio (LLR) value of zero to the punctured bits. See~\cite{NCL:13c} for details. 

The puncturing transform of a vector $\bv{x}$ of $M$ bits is also defined in two equivalent ways. In the first way, we define a vector $\lngp{\bv{x}}$ of length $N = 2^{\lceil \log_2 M \rceil}$ with indices $\mathcal{P}$ set to $\ourp$. We then copy $\bv{x}$ into $\lngp{\bv{x}}$ in order. That is, removing from $\lngp{\bv{x}}$ the indices in $\mathcal{P}$ recovers $\bv{x}$. Next, we compute $\lngp{\bv{u}}$, as explained in \Cref{subsec:PolarTransform}. We note that by the special choice of $\mathcal{P}$, we will never encounter a `$\emptyset$' entry in~\eqref{eq:OperationTables}. Lastly, we define $\bv{u}$ as the result of removing the first $N-M$ entries from $\lngp{\bv{u}}$.

Observe that had we replaced the entries in $\mathcal{P}$ with arbitrary binary numbers, the last $M$ entries of $\lngp{\bv{u}}$ would have been the same as the construction above. This is not surprising, since the generator matrix of the seminal polar codes is upper-triangular, after we apply bit reversal to the columns. This is the second way of defining the puncturing transform: replace every $\ourp$ with an arbitrary bit.

\begin{remark} \label{rmk:usuffix}
	Note that $\bv{u}$ equals the suffix of length $M$ of $\lngp{\bv{u}}$. That is, for $0 \leq i < M$, $u_i = \lngp{u}_{i+|\mathcal{P}|} = \lngp{u}_{i+N-M}$.
\end{remark}

\section{The `Inferior' and `Improved' Relations}
In this section, we define the `inferior' and `improved' relations between two joint distributions. Throughout, let $A(x_0;y_0)$ and $B(x_1;y_1)$ be joint distributions over $\mathcal{X} \times \mathcal{Y}_0$ and $\mathcal{X} \times \mathcal{Y}_1$, respectively. We denote that $A$ is inferior to $B$ by $A \inferior B$ and that $A$ is improved from $B$ by $A \improved B$. In fact, we only need to specify when $A \inferior B$ holds, since $A \inferior B$ if and only if $B \improved A$.

To define the `inferior' relation, we define two auxiliary relations between joint distributions.

\begin{itemize}
	\item \textbf{Degradation:} We say that $A$ is (stochastically) degraded from $B$, denoted $A \degraded B$, if there exists a conditional distribution $Q(y_0|y_1)$ over $\mathcal{Y}_0 \times \mathcal{Y}_1$ such that
		\begin{equation} \label{eq:degraded}
			A(x_0;y_0) = \sum_{y_1} B(x_0;y_1) Q(y_0|y_1) .
		\end{equation}
	\item \textbf{Input Permutation:} We say that $A$ has undergone an input permutation, resulting in  $A'$ if there exists a function $f: \mathcal{Y}_0 \to \mathcal{X}$ such that
		\begin{equation} \label{eq:permutation}
			A'(x_0;y_0) = A(x_0 \oplus f(y_0);y_0) .
		\end{equation}
		We denote this by $A' \permutation A$. Note that, like $A$, $A'$ is defined over $\mathcal{X} \times \mathcal{Y}_0$.
\end{itemize}

We now define that $A \inferior B$ if we can identify a finite sequence of `degradation' and `input permutation' relations that will lead to $A$ from $B$. In other words, there exists $0 < t < \infty$, a sequence of joint distributions $C_1,C_2,\ldots,C_{t-1}$, and a sequence $\text{r}_1,\text{r}_2,\ldots,\text{r}_t \in \{ \text{d}, \text{p} \}$ such that
\begin{equation} \label{eq:chain}
	A \joker{1} C_1 \joker{2} C_2 \joker{3} \cdots \joker{t-1} C_{t-1} \joker{t} B .
\end{equation}
Note that, essentially by definition, $\inferior$ is a transitive relation.
\subsection{Order Preservation}
For a joint distribution $A(x_0;y_0)$, we denote by $Z(A), K(A), H(A)$ the Bhattacharyya parameter $Z(X_0|Y_0)$, the total variation distance $K(X_0|Y_0)$, and the conditional entropy $H(X_0|Y_0)$, respectively, where $(X_0,Y_0)$ are distributed according to $A$. It is well known that if $A \degraded B$, then $Z(A) \geq Z(B)$, $K(A) \leq K(B)$, and $H(A) \geq H(B)$. The following lemma asserts that these inequalities also hold for $\inferior$.
\begin{lemma} \label{lem:inferiorYieldsZKH}
If $A \inferior B$, then $Z(A) \geq Z(B)$, $K(A) \leq K(B)$, and $H(A) \geq H(B)$.
\end{lemma}
\begin{IEEEproof}
	By definition of $\inferior$, and since the assertion in the lemma holds when $\inferior$ is replaced by $\degraded$, it suffices to show that it holds when $\inferior$ is replaced by $\permutation$. This follows easily.
\end{IEEEproof}

It is also well known that both $\boxast$ and $\circledast$ preserve $\degraded$. The following lemma generalizes this to $\inferior$.
\begin{lemma} \label{lem:inferiorYieldsMinusPlusOrdering}
	Let $A' \inferior A$ and $B' \inferior B$, then %
	\[
		A' \boxast B' \inferior A \boxast B \quad \text{and} \quad A' \circledast B' \inferior A \circledast B.
	\]	
\end{lemma}
\begin{IEEEproof}
See Appendix.
\end{IEEEproof}

The following lemma gives credence to names `superb' and `pitiful' for $\ourS$ and $\ourP$. Namely, it shows that $\ourS$ is `improved' with respect to all other distributions while $\ourP$ is `inferior' to all other distributions.
\begin{lemma} \label{lem:AbetwennPandS}
	Let $A(x_0;y_0)$ be a joint distribution over $\mathcal{X} \times \mathcal{Y}_0$. Then,
	\[
\ourP \inferior A \inferior \ourS .
	\]
\end{lemma}
\begin{IEEEproof} See Appendix.
\end{IEEEproof}

\subsection{The Equivalence Relation and Resulting Simplifications}
If $A \inferior B$ and $B \inferior A$, we denote $A \equiv B$ and call this the `equivalence' relation.

Above, we defined the special distributions $\ourS$ and $\ourP$. In the shortened (punctured) transform, these distributions replace the distribution $W(x;y)$ in the indices $\mathcal{S}$ ($\mathcal{P}$). Hence, they will take part in `$-$' and `$+$' operations (`$\boxast$' and `$\circledast$'). 
The following lemma shows that the results of such transforms involving $\ourS$ and $\ourP$ can be simplified using the equivalence relation. 

\begin{lemma} \label{lem:equivalenceRelationsForSandP}
	Let $A$ and $B$ be joint distributions. The following table summarizes the results of applying $\boxast$ and $\circledast$ operations to combinations of $A$, $B$, $\ourS$, and $\ourP$, up to equivalence.

\begin{equation}\label{eq:OperationTablesDistribution}
	\begin{tabular}{ c| c  c  c }
$\boxast$ & $B$           & $\ourS$ & $\ourP$ \\
\hline
$A$       & $A\boxast B $ & $A$     & $\ourP$ \\ 
$\ourS$   & $B$           & $\ourS$ & $\ourP$ \\ 
$\ourP$   & $\ourP$       & $\ourP$  & $\ourP$  
\end{tabular} \quad ;  \quad
\begin{tabular}{ c| c  c  c }
$\circledast$ & $B$               & $\ourS$ & $\ourP$ \\
\hline
$A$           & $A\circledast B $ & $\ourS$ & $A$ \\ 
$\ourS$       & $\ourS$           & $\ourS$ & $\ourS$ \\ 
$\ourP$       & $B$               & $\ourS$ & $\ourP$  
\end{tabular}.
\end{equation}

\end{lemma}
\begin{IEEEproof}
	See Appendix.
\end{IEEEproof}

\begin{remark} \label{rem:connectionBetweenTables} Tables~\eqref{eq:OperationTables} and \eqref{eq:OperationTablesDistribution} are connected by substitution. Namely, if in \eqref{eq:OperationTablesDistribution} we replace $\ourS$, $\ourP$, $\boxast$, $\circledast$ with $\ours$, $\ourp$, $\oplus$, $\triangleright$, then it is consistent with \eqref{eq:OperationTables}, if we now think of $A$ and $B$ as bits.
\end{remark}

\begin{remark} \label{rem:SandPConsistentwithsandp}
	The distribution $\ourS$ ($\ourP$) is consistent with the second way of defining the shortening (puncturing) transform. Namely, consider the pair of random vectors $\bv{X},\bv{Y}$ of length $M$, drawn i.i.d. according to $W(x;y)$.
	\begin{itemize}
		\item \emph{Shortening}: Let $\lngs{\bv{X}}$ be the random vector defined in the second way of shortening. By definition, all entries $\lngs{X}_i$ for $i \in \mathcal{S}$ are $0$ with probability $1$. Also, since for $i \in \mathcal{S}$ we do not transmit the corresponding symbol over the channel, $\lngs{Y}_i = ?$. Thus, pairs $(\lngs{X}_i,\lngs{Y}_i)$ for $i \in \mathcal{S}$ are distributed according to $\ourS$. As a consequence of this and \Cref{rmk:uprefix}, for $0 \leq i < M$,
			\begin{IEEEeqnarray}{rCl}
				Z(U_i|U^{i-1},\bv{Y}) &=& Z(\lngs{U}_i|\lngs{U}^{i-1},\lngs{\bv{Y}}), \label{eq:shortening:Zequality} \\
				K(U_i|U^{i-1},\bv{Y}) &=& K(\lngs{U}_i|\lngs{U}^{i-1},\lngs{\bv{Y}}). \label{eq:shortening:Kequality}
			\end{IEEEeqnarray}
			
		\item \emph{Puncturing}: Let $\lngp{\bv{X}}$ be the random vector defined in the second way of puncturing. By definition, we do not care about the value nor the distribution of any entry $\lngp{X}_i$ for $i \in \mathcal{P}$. However, we find it useful to set their distribution to be uniform and i.i.d. Also, since for $i \in \mathcal{P}$ we do not transmit the corresponding symbol over the channel, $\lngp{Y}_i = ?$. Thus, pairs $(\lngp{X}_i,\lngp{Y}_i)$ for $i \in \mathcal{P}$ are distributed according to $\ourP$. The reason for this choice is that now $\lngp{U}_0^{N-M-1}$ is independent of the triplet $\lngp{U}_{N-M}^N = \bv{U}$, $\bv{X}$, and $\bv{Y}$. This follows from the observation at the end of \Cref{subsec:puncturingTransform}.
Thus, for $0 \leq i < M$,
			\begin{IEEEeqnarray}{rCl}
				Z(U_i|U^{i-1},\bv{Y}) &=& Z(\lngp{U}_{i+M}|\lngp{U}^{i+M-1},\lngp{\bv{Y}}), \label{eq:puncturing:Zequality} \\
				K(U_i|U^{i-1},\bv{Y}) &=& K(\lngp{U}_{i+M}|\lngp{U}^{i+M-1},\lngp{\bv{Y}}). \label{eq:puncturing:Kequality}
			\end{IEEEeqnarray}

	\end{itemize}
\end{remark}
\section{Main Theorem}
The following theorem is the more general form of \Cref{thm:main}. Indeed, \Cref{thm:main} is a special case of \Cref{thm:mainAndGeneral}, where we obtain \eqref{eq:mainTheorem:K} from \eqref{eq:mainTheoremGeneralized:K} by defining $W(x;y)$ as being over $\mathcal{X} \times \{?\}$.
\begin{theorem} \label{thm:mainAndGeneral}
	Let $W(x;y)$ be a joint distribution over $\mathcal{X} \times \mathcal{Y}$. Let $\bv{X},\bv{Y}$ be a pair of random vectors of length $M$, with each $(X_i,Y_i)$ sampled independently from $W$. Let $\bv{U}$ of length $M$ be the result of transforming $\bv{X}$ via either the shortening transform or the puncturing transform. Fix $0 < \beta < 1/2$ and $\epsilon > 0$. Then, there exists $M_0$ such that for \underline{all} $M \geq M_0$,
\begin{IEEEeqnarray}{rCl}
	\frac{1}{M}\left|\left\{ i : Z(U_i|U^{i-1},\bv{Y}) < 2^{-M^\beta} \right\}\right| & > & 1 - H(X|Y) - \epsilon , \IEEEeqnarraynumspace \label{eq:mainTheoremGeneralized:Z}\\
	\frac{1}{M}\left|\left\{ i : K(U_i|U^{i-1},\bv{Y}) < 2^{-M^\beta} \right\}\right| & > & H(X|Y) - \epsilon. \label{eq:mainTheoremGeneralized:K}
\end{IEEEeqnarray}
\end{theorem}

The proof will be divided into two conceptual stages. In the first, we limit $M$ to be of a special form. That is, for some fixed $t$, $M = a \cdot 2^{n-t}$, where $a \in \{2^{t-1}+1, 2^{t-1}+2,\ldots, 2^{t}\}$. In the second stage, we show that such a restriction is not necessary.

The first stage is given in the following lemma.
\begin{lemma} \label{lemm:Mconstrained}
	Let $W(x;y)$, $\bv{X}$, $\bv{Y}$, and $\bv{U}$ be as in \Cref{thm:mainAndGeneral}. Fix $0 < \beta' < 1/2$ and $\epsilon' > 0$. Fix integers $t > 0$ and $a \in \{2^{t-1}+~1, 2^{t-1}+~2,\ldots, 2^{t}\}$. There exists $n_0$ such that for \underline{all} $n \geq n_0$, if $M = a \cdot 2^{n-t}$, then for $N = 2^n$,
\begin{IEEEeqnarray}{rCl}
	\frac{1}{M}\left|\left\{ i : Z(U_i|U^{i-1},\bv{Y}) < 2^{-N^{\beta'}} \right\}\right| & > & 1 - H(X|Y) - \epsilon' , \label{eq:lemm:Mconstrained:Z} \IEEEeqnarraynumspace\\
	\frac{1}{M}\left|\left\{ i : K(U_i|U^{i-1},\bv{Y}) < 2^{-N^{\beta'}} \right\}\right| & > & H(X|Y) - \epsilon'. \label{eq:lemm:Mconstrained:K}
\end{IEEEeqnarray}
\end{lemma}

Observe that in \eqref{eq:lemm:Mconstrained:Z} and \eqref{eq:lemm:Mconstrained:K}, the inequality is given in terms of $N \geq M$ in the exponential, and thus is stronger than had it been given in terms of $M$, as is done in \Cref{thm:mainAndGeneral}.
\begin{IEEEproof}
	The proofs for the shortening case and the puncturing case are similar. We show here in detail the proof for the shortening case.
	First, note that $n$ is consistent with the first equality in~\eqref{eq:n}, and indeed $N=2^n$ as in~\eqref{eq:N}.
	For $0 \leq i < N$, define the joint distribution $A_i$ as
	\[
		A_i =
		\begin{cases}
			W, & i \notin \mathcal{S}, \\ 
				\ourS , & i \in \mathcal{S}. 
		\end{cases}
\]
Note that by our choice of $\mathcal{S}$ in \eqref{eq:Sset} and the special structure $M = a \cdot 2^{n-t}$, the vector of joint distributions
$\begin{bmatrix} A_0 &  A_1 & \cdots & A_{N-1} \end{bmatrix}$ has period $2^t$. Indeed, consider the subvector $\begin{bmatrix} A_{2^t \cdot k} & A_{2^t \cdot k + 1} & \cdots & A_{2^t \cdot k + 2^t - 1} \end{bmatrix}$, for $0 \leq k < 2^{n-t}$. When bit reversing its entries, we get 
\begin{equation} \label{eq:WSperiod}
	\begin{bmatrix} \smash[b]{\underbrace{\begin{matrix}W & W & \cdots & W\end{matrix}}_{a}} & 
\smash[b]{\underbrace{\begin{matrix}\ourS & \ourS & \cdots & \ourS\end{matrix}}_{2^t-a}} \end{bmatrix}. \vphantom{\smash[t]{\underbrace{\begin{matrix}W \end{matrix}}_{a}}} 
\end{equation}
		   As a consequence, for any $\bv{b}_{(t)} = \begin{bmatrix} b_0 & b_1 & \cdots & b_{t-1} \end{bmatrix} \in \{0,1\}^t$, all the entries of
	\[
		\begin{bmatrix} A_0 &  A_1 & \cdots & A_{N-1} \end{bmatrix}^{[\bv{b}_{(t)}]}
	\]
are equal, i.e., the same joint distribution. Denote this distribution by $\Omega_{\bv{b}_{(t)}}$. Observe from \eqref{eq:WSperiod} that the mean conditional entropy of all such $\Omega_{\bv{b}_{(t)}}$ is $(a \cdot H(X|Y) + (2^t -a) \cdot 0)/2^t = a \cdot 2^{-t} \cdot H(X|Y) = M/N \cdot H(X|Y)$.

We are now in the scenario of identical distributions, undergoing a seminal polar transform of depth $n-t$. Calling upon standard results in polar codes\footnote{The inequality on $Z$ is given in \cite{ArikanTelatar:09c}, while for $K$ we can, for example, combine \cite[Prop. 4]{ShuvalTal:19.2p} with 
\cite[Lemma 2]{Tal:17.2p}.}, there exists an $n_0$ such that for all $n > n_0$,
\begin{IEEEeqnarray*}{rCl}
	\frac{1}{N}\left|\left\{ i : Z(\lngs{U}_i|\lngs{U}^{i-1},\lngs{\bv{Y}}) < 2^{-\left(\frac{N}{2^{t}}\right)^{\beta''}} \right\}\right| & > & 1 - \frac{M}{N} H(X|Y) - \epsilon'' , \\[0.1cm]
	\frac{1}{N}\left|\left\{ i : K(\lngs{U}_i|\lngs{U}^{i-1},\lngs{\bv{Y}}) < 2^{-\left(\frac{N}{2^{t}}\right)^{\beta''}} \right\}\right| & > & \frac{M}{N} H(X|Y) - \epsilon'', 
\end{IEEEeqnarray*}
where $\epsilon'' = \epsilon'/2$ and $\beta'' = \frac{\beta' + \frac{1}{2}}{2}$. Note that $\epsilon'' < \epsilon' \cdot M/N$.%

Recall that in the first way of describing the shortening transform, the last $N-M$ entries of $\lngs{\bv{u}}$ are all $\ours$. Thus, the joint distributions $(\lngs{U}_i;\lngs{U}^{i-1},\lngs{\bv{Y}})$, where $M \leq i < N$, are all equivalent to $\ourS$, by \Cref{rem:connectionBetweenTables,rem:SandPConsistentwithsandp}. Hence, for $M \leq i < N$, $Z(\lngs{U}_i|\lngs{U}^{i-1},\lngs{\bv{Y}}) = 0$ and $K(\lngs{U}_i|\lngs{U}^{i-1},\lngs{\bv{Y}}) = 1$. Therefore, if we limit $i$ in the braces to $0 \leq i < M$, recall that $\lngs{Y}_{N-M}^N=?? \cdots ?$, and use~\Cref{rmk:uprefix}, we obtain
\begin{IEEEeqnarray*}{l}
											     \frac{1}{N}\left|\left\{ 0 \leq i < M : Z(U_i|U^{i-1},\bv{Y}) < 2^{-\left(\frac{N}{2^{t}}\right)^{\beta''}} \right\}\right|  \\ 
											     \qquad\qquad  \qquad \qquad\qquad  \qquad\qquad  > \frac{M}{N} - \frac{M}{N} H(X|Y) - \epsilon'' , \\
											     \frac{1}{N}\left|\left\{0\leq i < M : K(U_i|U^{i-1},\bv{Y}) < 2^{-\left(\frac{N}{2^{t}}\right)^{\beta''}} \right\}\right|  \\ 
											     \qquad \qquad\qquad\qquad   \qquad\qquad \qquad   > \frac{M}{N} H(X|Y) - \epsilon''.
\end{IEEEeqnarray*}
Multiplying both sides by $N/M$ and further requiring that $n_0$ be large enough so that $(N/2^t)^{\beta''} > N^{\beta'}$, which is possible as $\beta'' > \beta'$, completes the proof for the shortening case.

In the puncturing case, we apply the above mechanics, extending $\bv{U}$ and $\bv{Y}$ to $\lngp{\bv{U}}$ and $\lngp{\bv{Y}}$, respectively. We then need to consider only the suffix of $\lngp{\bv{U}}$, due to~\Cref{rem:SandPConsistentwithsandp}. 
\end{IEEEproof}

The following corollary strengthens \Cref{lemm:Mconstrained} by setting a single $n_0$ that holds for all $a \in \{2^{t-1}, 2^{t-1}+1,\ldots, 2^{t}\}$.  Here the range of $a$ is extended to also contain $2^{t-1}$. Note that $n$ and $N=2^n$ are not consistent with \eqref{eq:N} and \eqref{eq:n} for $a = 2^{t-1}$.
\begin{corollary} \label{cor:lemm:Mconstrained}
	Let $W(x;y)$, $\bv{X}$, $\bv{Y}$, and $\bv{U}$ be as in \Cref{thm:mainAndGeneral}. Fix $0 < \beta' < 1/2$, $\epsilon' > 0$ and $t > 0$. There exists $n_0$ such that for \underline{all} $n \geq n_0$, if $M = a \cdot 2^{n-t}$, where $a \in \{2^{t-1}, 2^{t-1}+1,\ldots, 2^{t}\}$, then for $N = 2^n$,
	\eqref{eq:lemm:Mconstrained:Z} and \eqref{eq:lemm:Mconstrained:K} hold.
\end{corollary}

\begin{IEEEproof}
	For each $a \in \{2^{t-1}+1, 2^{t-1}+2,\ldots, 2^{t}\}$, \Cref{lemm:Mconstrained} holds for some $n_0$. 
	For the case $a = 2^{t-1}$, take $\beta'' = \frac{\beta' + \frac{1}{2}}{2}$ and use \Cref{lemm:Mconstrained} with $t=1$ to show that \eqref{eq:lemm:Mconstrained:Z} and \eqref{eq:lemm:Mconstrained:K} hold with $N/2$ and $\beta''$ in place of $N$ and $\beta'$, respectively, for some $n_0$. Now take the largest $n_0$ and further require that it is large enough so that $(N/2)^{\beta''} > N^{\beta'}$.
\end{IEEEproof}

\begin{IEEEproof}[Proof of \Cref{thm:mainAndGeneral}]
	We focus here on the shortening case.  Take $\epsilon' = \epsilon/2$, $\beta' = \beta$ and set $t$ such that $2^{1-t} < \epsilon'$. Let $n_0$ be as in \Cref{cor:lemm:Mconstrained}. We claim that $M_0 = 2^{n_0}$. Denote 
	\begin{IEEEeqnarray*}{rClCrCl}
		\chop{a} &=& \left\lfloor \frac{M}{2^{n-t}} \right\rfloor, &\quad& \chop{M} &=& \chop{a}\cdot 2^{n-t}, \\[0.1cm]
		\extend{a} &=& \left\lceil \frac{M}{2^{n-t}} \right\rceil, &\quad& \extend{M} &=& \extend{a}\cdot 2^{n-t}.
	\end{IEEEeqnarray*}
	Observe that both $\chop{M}$ and $\extend{M}$ are of the form $a\cdot2^{n-t}$ with $a \in \{2^{t-1}, 2^{t-1}+1,\ldots, 2^{t}\}$. These are the tightest choices of this form such that $N/2 \leq \chop{M} \leq M \leq \extend{M}$. Moreover, $\extend{a}-\chop{a} \leq 1$, yielding $\extend{M}- M \leq 2^{n-t}$ and $M-\chop{M} \leq 2^{n-t}$.

	We first prove \eqref{eq:mainTheoremGeneralized:Z}. For this, we consider the random vectors $\bv{U}, \lngs{\bv{U}}, \bv{X}, \lngs{\bv{X}}, \bv{Y}, \lngs{\bv{Y}}$ for our case of interest, i.e., shortening from length $N$ to length $M$. We will also consider the corresponding vectors for the case of shortening the same $N$ to length $\extend{M}$, denoted $\extend{\bv{U}}, \extend{\lngs{\bv{U}}}, \extend{\bv{X}}, \extend{\lngs{\bv{X}}}, \extend{\bv{Y}}, \extend{\lngs{\bv{Y}}}$. 
	
	By \Cref{cor:lemm:Mconstrained}, \eqref{eq:lemm:Mconstrained:Z} holds for $\extend{M}$. Thus,
	\begin{IEEEeqnarray*}{cl}
		& 1 - H(X|Y)-\epsilon' \\ 
		& \quad < \frac{1}{\extend{M}}\left|\left\{0 \leq i < \extend{M} : Z(\extend{U}_i|\extend{U}^{i-1},\extend{\bv{Y}}) < 2^{-N^{\beta'}} \right\}\right|   \\
		& \quad \eqann[\leq]{a} \frac{1}{\extend{M}}\left|\left\{0 \leq i < M  : Z(\extend{U}_i|\extend{U}^{i-1},\extend{\bv{Y}}) < 2^{-N^{\beta'}} \right\}\right|  + \frac{\extend{M}-M}{\extend{M}} \\ 
		& \quad \eqann[\leq]{b} \frac{1}{\extend{M}}\left|\left\{0 \leq i < M  : Z(U_i|U^{i-1},\bv{Y}) < 2^{-N^{\beta'}} \right\}\right|  + \frac{\extend{M}-M}{\extend{M}} \\ 
		& \quad \eqann[\leq]{c} \frac{1}{M}\left|\left\{0 \leq i < M  : Z(U_i|U^{i-1},\bv{Y}) < 2^{-N^{\beta'}} \right\}\right|  + \frac{\extend{M}-M}{N/2} \\ 
		& \quad \eqann[\leq]{d} \frac{1}{M}\left|\left\{0 \leq i < M  : Z(U_i|U^{i-1},\bv{Y}) < 2^{-N^{\beta'}} \right\}\right|  + 2^{1-t} \\ 
		& \quad \eqann[<]{e} \frac{1}{M}\left|\left\{0 \leq i < M  : Z(U_i|U^{i-1},\bv{Y}) < 2^{-N^{\beta'}} \right\}\right|  + \epsilon'. 
	\end{IEEEeqnarray*}
	Rearranging and recalling that $\beta'=\beta$ yields~\eqref{eq:mainTheoremGeneralized:Z}. We now explain inequalities \eqannref{a}--\eqannref{e}.
	\begin{itemize}
        \item \eqannref{a}: The index set on the right-hand-side is smaller, as $\extend{M} \geq M$. The contribution of the non-counted indices is at most $\extend{M}-M$. \vspace{0.1cm}
		\item \eqannref{b}: We have for $0 \leq i < M$ that
			\begin{IEEEeqnarray*}{rCl}
				Z(\extend{U}_i|\extend{U}^{i-1},\extend{\bv{Y}}) &=& Z(\extend{\lngs{U}}_i|\lngs{\extend{U}}^{i-1},\lngs{\extend{\bv{Y}}})  \\ &\leq& Z(\lngs{U}_i|\lngs{U}^{i-1},\lngs{\bv{Y}}) =  Z(U_i|U^{i-1},\bv{Y}),
			\end{IEEEeqnarray*}
			where the equalities follow from \eqref{eq:shortening:Zequality} and the inequality follows from \Cref{lem:inferiorYieldsZKH} as the joint distribution of $(\lngs{U}_i; \lngs{U}^{i-1},\lngs{\bv{Y}})$ is improved from $(\extend{\lngs{U}}_i; \lngs{\extend{U}}^{i-1},\lngs{\extend{\bv{Y}}})$. Indeed, this latter observation follows from \Cref{lem:inferiorYieldsMinusPlusOrdering,lem:AbetwennPandS}.
		\item \eqannref{c}: This follows from $M \leq \extend{M}$ and $N/2 \leq \extend{M}$.
		\item \eqannref{d}: This is due to $\extend{M}-M \leq 2^{n-t}$ and $N=2^n$.
		\item \eqannref{e}: We defined $t$ such that $2^{1-t}< \epsilon'$. 
	\end{itemize}

	We now prove \eqref{eq:mainTheoremGeneralized:K}. For this, we again consider the random vectors $\bv{U}, \lngs{\bv{U}}, \bv{X}, \lngs{\bv{X}}, \bv{Y}, \lngs{\bv{Y}}$ for our case of interest, i.e., shortening from length $N$ to length $M$. We further consider the corresponding vectors for the case of shortening the same $N$ to length $\chop{M}$, denoted $\chop{\bv{U}}, \chop{\lngs{\bv{U}}}, \chop{\bv{X}}, \chop{\lngs{\bv{X}}}, \chop{\bv{Y}}, \chop{\lngs{\bv{Y}}}$. 
	
	By \Cref{cor:lemm:Mconstrained}, \eqref{eq:lemm:Mconstrained:K} holds for $\chop{M}$. Thus,
	\begin{IEEEeqnarray*}{cl}
		& H(X|Y)-\epsilon' \\ 
		& \quad < \frac{1}{\chop{M}}\left|\left\{0 \leq i < \chop{M} : K(\chop{U}_i|\chop{U}^{i-1},\chop{\bv{Y}}) < 2^{-N^{\beta'}} \right\}\right|   \\
		& \quad \eqann[\leq]{a} \frac{1}{\chop{M}}\left|\left\{0 \leq i < \chop{M} : K(U_i|U^{i-1},\bv{Y}) < 2^{-N^{\beta'}} \right\}\right|   \\
		& \quad \eqann[\leq]{b} \frac{1}{\chop{M}}\left|\left\{0 \leq i < M : K(U_i|U^{i-1},\bv{Y}) < 2^{-N^{\beta'}} \right\}\right|   \\
		& \quad = \left(\frac{1}{M}+\frac{1}{\chop{M}}-\frac{1}{M}\right)\left|\left\{0 \leq i < M : K(U_i|U^{i-1},\bv{Y}) < 2^{-N^{\beta'}} \right\}\right|   \\
		& \quad \eqann[\leq]{c} \frac{1}{M}\left|\left\{0 \leq i < M : K(U_i|U^{i-1},\bv{Y}) < 2^{-N^{\beta'}} \right\}\right| + \frac{M-\chop{M}}{\chop{M}}  \\
		& \quad \eqann[\leq]{d} \frac{1}{M}\left|\left\{0 \leq i < M : K(U_i|U^{i-1},\bv{Y}) < 2^{-N^{\beta'}} \right\}\right| + \frac{M-\chop{M}}{N/2}  \\
		& \quad \eqann[\leq]{e} \frac{1}{M}\left|\left\{0 \leq i < M : K(U_i|U^{i-1},\bv{Y}) < 2^{-N^{\beta'}} \right\}\right| + 2^{1-t}  \\
		& \quad \eqann[<]{f} \frac{1}{M}\left|\left\{0 \leq i < M  : K(U_i|U^{i-1},\bv{Y}) < 2^{-N^{\beta'}} \right\}\right|  + \epsilon'. 
	\end{IEEEeqnarray*}
	Rearranging yields~\eqref{eq:mainTheoremGeneralized:K}. We now explain inequalities \eqannref{a}--\eqannref{f}.
	\begin{itemize}
		\item \eqannref{a}: We have for $0 \leq i < \chop{M}$ that
            \vspace{-0.1cm}\begin{IEEEeqnarray*}{rCl}
				K(\chop{U}_i|\chop{U}^{i-1},\chop{\bv{Y}}) &=& K(\chop{\lngs{U}}_i|\lngs{\chop{U}}^{i-1},\lngs{\chop{\bv{Y}}})  \\ &\leq& K(\lngs{U}_i|\lngs{U}^{i-1},\lngs{\bv{Y}}) =  K(U_i|U^{i-1},\bv{Y}),
			\end{IEEEeqnarray*}
			where the equalities follow from \eqref{eq:shortening:Kequality} and the inequality follows from \Cref{lem:inferiorYieldsZKH} as the joint distribution of $(\lngs{U}_i; \lngs{U}^{i-1},\lngs{\bv{Y}})$ is inferior to $(\chop{\lngs{U}}_i; \chop{\lngs{U}}^{i-1},\lngs{\chop{\bv{Y}}})$. Indeed, this latter observation follows from \Cref{lem:inferiorYieldsMinusPlusOrdering,lem:AbetwennPandS}.
		\item \eqannref{b}: As $M \geq \chop{M}$, the right-hand-side is the size of a larger set than the left-hand-side. 
		\item \eqannref{c}: The size of the set is at most $M$, and $M\cdot(1/\chop{M} - 1/M) = (M-\chop{M})/\chop{M}$. 
		\item \eqannref{d}: This is due to $\chop{M}\leq N/2$. 
		\item \eqannref{d}: This is due to $M-\chop{M} \leq 2^{n-t}$ and $N=2^n$.
		\item \eqannref{f}: We defined $t$ such that $2^{1-t}< \epsilon'$. 
	\end{itemize}
This completes the proof for the shortening case. 

The puncturing case uses similar mechanics. The proof of~\eqref{eq:mainTheoremGeneralized:Z} for puncturing follows along the lines of the proof of~\eqref{eq:mainTheoremGeneralized:K} for shortening. The proof of~\eqref{eq:mainTheoremGeneralized:K} for puncturing follows along the lines of the proof of~\eqref{eq:mainTheoremGeneralized:Z} for shortening. %
\end{IEEEproof} 


\begin{appendix}
\begin{IEEEproof}[Proof of \Cref{lem:inferiorYieldsMinusPlusOrdering}]
	Since $A' \inferior A$ and $B' \inferior B$, we recall \eqref{eq:chain} and denote
	\[
		A' \jokerA{1} C_1^A \jokerA{2} C_2^A \jokerA{3} \cdots \jokerA{t_A - 1} C_{t_A - 1}^A \jokerA{t_A} A  
	\]
	and
	\[
		B' \jokerB{1} C_1^B \jokerB{2} C_2^B \jokerB{3} \cdots \jokerA{t_B - 1} C_{t_B - 1}^B \jokerB{t_B} B,
	\]
	where $\text{r}_1^A,\text{r}_2^A,\ldots, \text{r}_{t_A}^A \in \{ \text{d}, \text{p} \}$ and also $\text{r}_1^B,\text{r}_2^B,\ldots, \text{r}_{t_B}^B \in \{ \text{d}, \text{p} \}$. The proof is by induction on $t_A + t_B$.

	For the base case, take $t_A + t_B = 0$. That is, $t_A = t_B = 0$, which implies that $A' = A$ and $B' = B$, and there is nothing to prove. 

	For the induction step, assume the claim holds when $t_A + t_B = \ell$, and consider a case where $t_A + t_B = \ell + 1$. Thus, either $t_A > 0$ or $t_B > 0$ (or both). If $t_A > 0$, it suffices to prove that
	\[
		A' \boxast B' \inferior C_1^A \boxast B' \quad \text{and} \quad A' \circledast B' \inferior C_1^A \circledast B',
	\]
	since we have by the induction hypothesis that
	\[
		C_1^A \boxast B'  \inferior A \boxast B \quad \text{and} \quad C_1^A \circledast B' \inferior A \circledast B,
	\]
	and the claim follows by the transitivity of the $\inferior$ relation. Similarly, if $t_B > 0$ it suffices to prove that
	\[
		A' \boxast B' \inferior A' \boxast C_1^B \quad \text{and} \quad A' \circledast B' \inferior A' \circledast C_1^B.
	\]

	There are 8 cases to consider, since there are two options for the transform, $\boxast$ and $\circledast$; two options for the gateway joint distribution, $C_1^A$ and $C_1^B$; and two options of getting to the gateway joint distribution, $\degraded$ and $\permutation$. The first 4 cases will deal with $C_1^A$ and the last 4 with $C_1^B$. In the interest of keeping the notation light, in the first 4 cases we rename $C_1^A$ to $A$ and $B'$ to $B$ and in the last 4 cases we rename $C_1^B$ to $B$ and $A'$ to $A$. The cases in which we consider $\degraded$ are brought here completeness, as they have already been proven in \cite[Lemma 4.7]{Korada:09z}.

	\begin{enumerate}
		\item We show that
			\[
A' \degraded A \Longrightarrow A' \boxast B \degraded A \boxast B.
			\]
			If $A' \degraded A$ then by \eqref{eq:degraded}, for some $Q(y_0'|y_0)$ we have
			\begin{equation} \label{eq:QforA}
				A'(x_0;y_0') = \sum_{y_0} A(x_0;y_0) Q(y_0'|y_0).
			\end{equation}
			Thus, by \eqref{eq:AboxastBdef},
			\begin{IEEEeqnarray*}{rCl}
				\IEEEeqnarraymulticol{3}{l}{(A' \boxast B)(u_0;y'_0,y_1')} \\
				\quad &=& \sum_{x_1} A'(u_0 \oplus x_1;y_0') B(x_1;y_1') \\
				      &=& \sum_{x_1} \sum_{y_0} A(u_0 \oplus x_1;y_0) Q(y_0'|y_0) B(x_1;y_1') \\
				      &=& \sum_{y_0} \sum_{x_1} A(u_0 \oplus x_1;y_0)  B(x_1;y_1') Q(y_0'|y_0) \\
				      &=& \sum_{y_0} (A \boxast B)(u_0;y_0,y_1') Q(y_0'|y_0).
			\end{IEEEeqnarray*}
			We now define
			\[
				Q'(y_0',y_1'|y_0,y_1) = \begin{cases}
					Q(y_0'|y_0), & y_1' = y_1 \\
					0, & \text{otherwise},
				\end{cases}
			\]
			and continue the above derivation as
			\begin{IEEEeqnarray*}{rCl}
				\IEEEeqnarraymulticol{3}{l}{\sum_{y_0} (A \boxast B)(u_0;y_0,y_1') Q(y_0'|y_0)} \\ 
				\quad &=& \sum_{y_0,y_1} (A \boxast B)(u_0;y_0,y_1) Q'(y_0',y_1'|y_0,y_1).
			\end{IEEEeqnarray*}
			The claim follows by \eqref{eq:degraded}.
			
		\item We show that
			\[
A' \degraded A \Longrightarrow A' \circledast B \degraded A \circledast B.
			\]
			As in the previous case, there exists $Q(y_0'|y_0)$ such that \eqref{eq:QforA} holds.
			Thus, by \eqref{eq:AcircledastBdef},
			\begin{IEEEeqnarray*}{rCl}
				\IEEEeqnarraymulticol{3}{l}{(A' \circledast B)(u_1;u_0',y'_0,y_1')} \\
				\quad &=& A'(u_0' \oplus u_1;y_0') B(u_1;y_1') \\
				      &=& \sum_{y_0} A(u_0' \oplus u_1;y_0) Q(y_0'|y_0) B(u_1;y_1') \\
				      &=& \sum_{y_0} A(u_0' \oplus u_1;y_0)  B(u_1;y_1') Q(y_0'|y_0) \\
				      &=& \sum_{y_0} (A \circledast B)(u_1;u_0',y_0,y_1') Q(y_0'|y_0).
			\end{IEEEeqnarray*}
			We now define
			\[
				Q'(u_0',y_0',y_1'|u_0,y_0,y_1) = \begin{cases}
					Q(y_0'|y_0), & y_1' = y_1, u_0' = u_0 \\
					0, & \text{otherwise},
				\end{cases}
			\]
			and continue the above derivation as
			\begin{IEEEeqnarray*}{rCl}
				\IEEEeqnarraymulticol{3}{l}{\sum_{y_0} (A \circledast B)(u_1;u_0',y_0,y_1') Q(y_0'|y_0)} \\ 
				\quad &=& \sum_{u_0,y_0,y_1} (A \circledast B)(u_1;u_0,y_0,y_1) Q'(u_0',y_0',y_1'|u_0,y_0,y_1).
			\end{IEEEeqnarray*}
			The claim follows by \eqref{eq:degraded}.

		\item We show that
			\[
A' \permutation A \Longrightarrow A' \boxast B \permutation A \boxast B.
			\]
			If $A' \permutation A$ then by \eqref{eq:permutation}, for some $f(y_0)$ we have
			\begin{equation} \label{eq:fforA}
				A'(x_0;y_0) = A(x_0 \oplus f(y_0) ;y_0).
			\end{equation}
			Thus, by \eqref{eq:AboxastBdef},
			\begin{IEEEeqnarray*}{rCl}
				\IEEEeqnarraymulticol{3}{l}{(A' \boxast B)(u_0;y_0,y_1)} \\
				\quad &=& \sum_{x_1} A'(u_0 \oplus x_1;y_0) B(x_1;y_1) \\
				      &=& \sum_{x_1} A(u_0 \oplus x_1 \oplus f(y_0);y_0) B(x_1;y_1) \\
				      &=& (A \boxast B)(u_0 \oplus f(y_0) ;y_0,y_1).
			\end{IEEEeqnarray*}
			We now define
			\[
				g(y_0,y_1) = f(y_0) 
			\]
			and continue the above derivation as
			\begin{IEEEeqnarray*}{rCl}
				\IEEEeqnarraymulticol{3}{l}{(A \boxast B)(u_0 \oplus f(y_0) ;y_0,y_1)} \\ 
				\quad &=& (A \boxast B)(u_0 \oplus g(y_0,y_1) ;y_0,y_1).
			\end{IEEEeqnarray*}
			The claim follows by \eqref{eq:permutation}.

		\item We show that
			\[
A' \permutation A \Longrightarrow A' \circledast B \inferior A \circledast B.
			\]
			Specifically, we show that
			\[
A' \permutation A \Longrightarrow A' \circledast B \degraded A \circledast B.
			\]
			As in the previous case, there exists $f(y_0)$ such that \eqref{eq:fforA} holds.
			Thus, by \eqref{eq:AcircledastBdef},
			\begin{IEEEeqnarray*}{rCl}
				\IEEEeqnarraymulticol{3}{l}{(A' \circledast B)(u_1;u_0',y_0',y_1')} \\
				\quad &=& A'(u_0' \oplus u_1;y_0') B(u_1;y_1') \\
				      &=& A(u_0' \oplus u_1 \oplus f(y_0');y_0') B(u_1;y_1') \\
				      &=& (A \circledast B)(u_1; u_0' \oplus f(y_0'), y_0',y_1').
			\end{IEEEeqnarray*}
			We now define
			\begin{multline*}
				Q'(u_0',y_0',y_1'|u_0,y_0,y_1) \\
				= \begin{cases}
					1, & (u_0' \oplus f(y_0'), y_0',y_1') = (u_0,y_0,y_1)  \\
					0, & \text{otherwise},
				\end{cases}
			\end{multline*}
			and continue the above derivation as
			\begin{IEEEeqnarray*}{rCl}
				\IEEEeqnarraymulticol{3}{l}{(A \circledast B)(u_1; u_0' \oplus f(y_0'), y_0,y_1')} \\ 
				\quad &=& \sum_{u_0,y_0,y_1} (A \circledast B)(u_1;u_0,y_0,y_1) Q'(u_0',y_0',y_1'|u_0,y_0,y_1).
			\end{IEEEeqnarray*}
			The claim follows by \eqref{eq:degraded}.

		\item We show that
			\[
B' \degraded B \Longrightarrow A \boxast B' \degraded A \boxast B.
			\]
			If $B' \degraded B$ then by \eqref{eq:degraded}, for some $Q(y_1'|y_1)$ we have
			\begin{equation} \label{eq:QforB}
				B'(x_1;y_1') = \sum_{y_1} B(x_1;y_1) Q(y_1'|y_1).
			\end{equation}
			Thus, by \eqref{eq:AboxastBdef},
			\begin{IEEEeqnarray*}{rCl}
				\IEEEeqnarraymulticol{3}{l}{(A \boxast B')(u_0;y'_0,y_1')} \\
				\quad &=& \sum_{x_1} A(u_0 \oplus x_1;y_0') B'(x_1;y_1') \\
				      &=& \sum_{x_1}  A(u_0 \oplus x_1;y_0') \sum_{y_1} B(x_1;y_1) Q(y_1'|y_1) \\
				      &=& \sum_{y_1} \sum_{x_1} A(u_0 \oplus x_1;y_0')  B(x_1;y_1) Q(y_1'|y_1)\\
				      &=& \sum_{y_1} (A \boxast B)(u_0;y_0',y_1) Q(y_1'|y_1) . 
			\end{IEEEeqnarray*}
			We now define
			\[
				Q'(y_0',y_1'|y_0,y_1) = \begin{cases}
					Q(y_1'|y_1), & y_0' = y_0 \\
					0, & \text{otherwise},
				\end{cases}
			\]
			and continue the above derivation as
			\begin{IEEEeqnarray*}{rCl}
				\IEEEeqnarraymulticol{3}{l}{\sum_{y_1} (A \boxast B)(u_0;y_0',y_1) Q(y_1'|y_1)} \\ 
				\quad &=& \sum_{y_0,y_1} (A \boxast B)(u_0;y_0,y_1) Q'(y_0',y_1'|y_0,y_1).
			\end{IEEEeqnarray*}
			The claim follows by \eqref{eq:degraded}.
			
		\item We show that
			\[
B' \degraded B \Longrightarrow A \circledast B' \degraded A \circledast B.
			\]
			As in the previous case, there exists $Q(y_1'|y_1)$ such that \eqref{eq:QforB} holds.

			Thus, by \eqref{eq:AcircledastBdef},
			\begin{IEEEeqnarray*}{rCl}
				\IEEEeqnarraymulticol{3}{l}{(A \circledast B')(u_1;u_0',y'_0,y_1')} \\
				\quad &=& A(u_0' \oplus u_1;y_0') B'(u_1;y_1') \\
				      &=&  A(u_0' \oplus u_1;y_0') \sum_{y_1} B(u_1;y_1) Q(y_1'|y_1) \\
				      &=& \sum_{y_1} A(u_0' \oplus u_1;y_0') B(u_1;y_1) Q(y_1'|y_1) \\
				      &=& \sum_{y_1} (A \circledast B)(u_1;u_0',y_0',y_1) Q(y_1'|y_1).
			\end{IEEEeqnarray*}
			We now define
			\[
				Q'(u_0',y_0',y_1'|u_0,y_0,y_1) = \begin{cases}
					Q(y_1'|y_1), & y_0' = y_0, u_0' = u_0 \\
					0, & \text{otherwise},
				\end{cases}
			\]
			and continue the above derivation as
			\begin{IEEEeqnarray*}{rCl}
				\IEEEeqnarraymulticol{3}{l}{\sum_{y_1} (A \circledast B)(u_1;u_0',y_0',y_1) Q(y_1'|y_1)} \\ 
				\quad &=& \sum_{u_0,y_0,y_1} (A \circledast B)(u_1;u_0,y_0,y_1) Q'(u_0',y_0',y_1'|u_0,y_0,y_1).
			\end{IEEEeqnarray*}
			The claim follows by \eqref{eq:degraded}.

		\item We show that
            \vspace{-0.1cm}\[
B' \permutation B \Longrightarrow A \boxast B' \permutation A \boxast B.
			\]
			If $B' \permutation B$ then by \eqref{eq:permutation}, for some $f(y_1)$ we have
			\begin{equation} \label{eq:fforB}
				B'(x_1;y_1) = B(x_1 \oplus f(y_1) ;y_1). 
			\end{equation}
			Thus, by \eqref{eq:AboxastBdef},
			\begin{IEEEeqnarray*}{rCl}
				\IEEEeqnarraymulticol{3}{l}{(A \boxast B')(u_0;y_0,y_1)} \\
				\quad &=& \sum_{x_1} A(u_0 \oplus x_1;y_0) B'(x_1;y_1) \\
				      &=& \sum_{x_1} A(u_0 \oplus x_1;y_0) B(x_1 \oplus f(y_1);y_1) \\
				      &\eqann{a}& \sum_{x_1} A(u_0 \oplus x_1 \oplus f(y_1);y_0) B(x_1;y_1) \\
				      &=& (A \boxast B)(u_0 \oplus f(y_1) ;y_0,y_1).
			\end{IEEEeqnarray*}
			Note that \eqannref{a} holds both when $f(y_1) = 0$ and $f(y_1) = 1$. In the former this is trivial and in the latter we're simply changing the order of summation.
			We now define
			\[
				g(y_0,y_1) = f(y_1) 
			\]
			and continue the above derivation as
            \vspace{-0.1cm}
			\begin{IEEEeqnarray*}{rCl}
				\IEEEeqnarraymulticol{3}{l}{(A \boxast B)(u_0 \oplus f(y_1) ;y_0,y_1)} \\ 
				\quad &=& (A \boxast B)(u_0 \oplus g(y_0,y_1) ;y_0,y_1).
			\end{IEEEeqnarray*}
			The claim follows by \eqref{eq:permutation}.
		\item We show that
            \vspace{-0.1cm}\[
B' \permutation B \Longrightarrow A \circledast B' \inferior A \circledast B.
			\]
			Specifically, we show that
            \vspace{-0.1cm}\[
B' \permutation B \Longrightarrow A \circledast B' \degraded C \permutation A \circledast B
			\]
			for a joint distribution $C$ we will shortly define.
			As in the previous case, there exists $f(y_1)$ such that \eqref{eq:fforB} holds.
            Further, let $f'(u_0,y_0,y_1) = f(y_1)$ and define
            \begin{IEEEeqnarray*}{rCl}
                C(u_1;u_0,y_0,y_1) &=& (A \circledast B)(u_1\oplus f'(u_0,y_0,y_1); u_0 , y_0,y_1) \\
                                   &=& (A \circledast B)(u_1\oplus f(y_1); u_0 , y_0,y_1).
\end{IEEEeqnarray*}
			Clearly, by \eqref{eq:permutation}, $C \permutation A \circledast B$.  Next, by \eqref{eq:AcircledastBdef},
			\begin{IEEEeqnarray*}{rCl}
				\IEEEeqnarraymulticol{3}{l}{(A \circledast B')(u_1;u_0',y_0',y_1')} \\
				\quad &=& A(u_0' \oplus u_1;y_0') B'( u_1; y_1') \\
				      &=& A(u_0' \oplus u_1;y_0') B( u_1 \oplus f(y_1'); y_1') \\
				      &=& A(u_0' \oplus f(y_1') \oplus u_1 \oplus f(y_1');y_0') B( u_1 \oplus f(y_1'); y_1') \\
				      &=& (A \circledast B)(u_1\oplus f(y_1'); u_0' \oplus f(y_1'), y_0',y_1') \\
				      &=& C(u_1; u_0' \oplus f(y_1'), y_0',y_1'). 
			\end{IEEEeqnarray*}
			We now define
			\begin{multline*}
				Q'(u_0',y_0',y_1'|u_0,y_0,y_1) \\
				= \begin{cases}
					1, & (u_0' \oplus f(y_1'), y_0',y_1') = (u_0,y_0,y_1)  \\
					0, & \text{otherwise},
				\end{cases}
			\end{multline*}
			and continue the above derivation as
			\begin{IEEEeqnarray*}{rCl}
				\IEEEeqnarraymulticol{3}{l}{C(u_1; u_0' \oplus f(y_1'), y_0',y_1')} \\ 
				\quad &=& \sum_{u_0,y_0,y_1} C(u_1;u_0,y_0,y_1) Q'(u_0',y_0',y_1'|u_0,y_0,y_1).
			\end{IEEEeqnarray*}
			The claim follows by \eqref{eq:degraded}.  \hspace{1em plus 1fill} \IEEEQEDhere
	\end{enumerate}
\end{IEEEproof}

	\begin{IEEEproof}[Proof of \Cref{lem:AbetwennPandS}]
	We first show that $\ourP \inferior A$. That is, we show that
	\[
\ourP \degraded C_1 \permutation C_2 \degraded A. 
	\]
	To this end, let $Q(y_0,x_1|y_0') = 1/2$ if $y_0'=y_0$ and $0$ otherwise. Then, by~\eqref{eq:degraded}, indeed $C_2 \degraded A$, with
\[
		C_2(x_0;y_0,x_1) = \sum_{y_0' \in \mathcal{Y}_0 } A(x_0;y_0') Q(y_0,x_1|y_0'). 
	\]
	Next, define $f(y_0, x_1) = x_1$. Then, by~\eqref{eq:permutation}, $C_1 \permutation C_2$ with
	\[ C_1(x_0 ; y_0, x_1) = C_2(x_0 \oplus f(y_0,x_1) ; y_0, x_1).\]  
	Observe by marginalization that $C_1(x_0) = 1/2$ for $x_0  \in \mathcal{X}$. 
	Finally, again by~\eqref{eq:degraded}, $\ourP \degraded C_1$ with 
	\[ 
	\ourP(x_0; y_0') = \sum_{y_0,x_1}C_1(x_0; y_0, x_1)Q'(y_0'|y_0,x_1), 
\] where $Q'(y_0'|y_0,x_1) = 1$ whenever $y_0' = ?$ and $0$ otherwise. 

Next, we show that $A \inferior \ourS$. That is, we show that 
\[ 
A \degraded D_1 \permutation D_2 \degraded \ourS. 
\]
First, let $R(y_0, x_0| y) = A(x_0 ; y_0)$. Then, by~\eqref{eq:degraded}, $D_2 \degraded \ourS$ with
\[ 
	D_2(x;x_0,y_0) = \sum_y \ourS(x;y)R(y_0,x_0 | y). 
\] 
Observe that $D_2(x;x_0,y_0) = A(x_0; y_0)$ when $x=0$ and $0$ otherwise.
Next, define $g(x_0,y_0) = x_0$. By~\eqref{eq:permutation}, $D_1 \permutation D_2$ with 
\[ D_1(x';x_0,y_0) = D_2(x'\oplus g(x_0, y_0) ; x_0, y_0).\]
Observe that 
\[ D_1(x'; x_0,y_0) = \begin{cases} A(x_0; y_0), & x' = x_0, \\ 0, & \text{otherwise}. \end{cases} \] 
For the final step, take $R'(y'|x_0,y_0) = 1$ if $y'=y_0$ and $0$ otherwise. Observe that %
\[
	A(x';y') = \sum_{x_0,y_0} D_1(x'; x_0, y_0)R'(y'|x_0,y_0).
\] 
Indeed, the sum is nonzero only when $x' = x_0$ and $y' = y_0$, in which case it equals $A(x';y')$. Thus, by~\eqref{eq:degraded}, $A \degraded D_1$.
\end{IEEEproof}
\begin{IEEEproof}[Proof of \Cref{lem:equivalenceRelationsForSandP}]
	Throughout the proof we will use~\eqref{eq:ourSDef} and~\eqref{eq:ourPDef}.

	\emph{Step 1: $A \boxast \ourS \equiv A$.} First, note by~\eqref{eq:AboxastBdef} that $(A \boxast \ourS)(u_0; y_0,?) = A(u_0;y_0)$. 
	To see the equivalence, we show that $A \inferior A \boxast \ourS \inferior A$. 
	Indeed, $A \boxast \ourS \degraded A$, since in \eqref{eq:degraded} we can take
\[
	Q(y_0,?|y') = \begin{cases} 1, & y' = y_0 ,\\
		0, & \text{otherwise},
	\end{cases}
\]
and $A \degraded A \boxast \ourS$, by \eqref{eq:degraded} with
\[
	Q(y'|y_0,?) = \begin{cases} 1, & y' = y_0 ,\\
		0, & \text{otherwise}.
	\end{cases}
\]

\emph{Step 2: $A \boxast \ourP \equiv \ourP$.} By \Cref{lem:AbetwennPandS}, $\ourP \inferior A \boxast \ourP$. It remains to show $A \boxast \ourP \inferior \ourP$. By~\eqref{eq:AboxastBdef}, $(A \boxast \ourP)(u_0; y_0,?) = A(y_0)/2$, where %
we denote $A(y_0) = A(0;y_0) + A(1;y_0)$. Next, $A \boxast \ourP \degraded \ourP$ by~\eqref{eq:degraded} with
$Q(y_0, ? | ? ) = A(y_0)$.

\emph{Step 3: $\ourS \boxast B \equiv B$.} By~\eqref{eq:AboxastBdef}, we have $(\ourS \boxast B)(u_0; ?,y_1) = B(u_0; y_1)$. Next, $\ourS \boxast B\equiv B$ by the same arguments as in Step 1.

\emph{Step 4: $\ourP \boxast B \equiv \ourP$.} By~\eqref{eq:AboxastBdef}, we have $(\ourP \boxast B)(u_0; ?,y_1) = B(y_1)/2$, where $B(y_1) = B(0;y_1)+B(y_1)$. Next, $\ourP \boxast B\equiv \ourP$ by the same arguments as in Step 2.

\emph{Step 5: $\ourS \boxast \ourS \equiv \ourS$, $\ourS \boxast \ourP \equiv \ourP$, $\ourP \boxast \ourS \equiv \ourP$, and $\ourP \boxast \ourP \equiv \ourP$.} These are special cases of Steps 1--4.

\emph{Step 6: $A \circledast \ourS \equiv \ourS$.} By \Cref{lem:AbetwennPandS}, $A \circledast \ourS \inferior \ourS$. It remains to show that $\ourS \inferior A \circledast \ourS$. By \eqref{eq:AcircledastBdef},
\[
	(A \circledast \ourS)(u_1;u_0,y_0,?) = \begin{cases}
		A(u_0;y_0), & u_1 = 0,\\
		0, & \text{otherwise}.
	\end{cases}
\]
Next, $\ourS \degraded A \circledast \ourS$ by~\eqref{eq:degraded} with
$Q(? | u_0,y_0, ? ) = 1$.

\emph{Step 7: $A \circledast \ourP \equiv A$.}
	First, note by~\eqref{eq:AcircledastBdef} that 
	\begin{equation}\label{eq:step7AP}
		(A \circledast \ourP)(u_1; u_0, y_0,?) = \frac{1}{2}A(u_0\oplus u_1;y_0).
	\end{equation}
	To see the equivalence, we show that $A \inferior A \circledast \ourP \inferior A$. 

	To show that $A \circledast \ourP \inferior A$ we show that
	\[
A \circledast \ourP \permutation C_1 \degraded A. 
\]
	To this end, for $u_0 \in \mathcal{X}$, let $Q(u_0,y_0,?|y_1) = 1/2$ if $y_1=y_0$ and $0$ otherwise. Then, by~\eqref{eq:degraded}, indeed $C_1 \degraded A$, with
\[
	C_1(u_1;u_0,y_0,?) = \sol{\sum_{y_1}} A(u_1;y_1) Q(u_0,y_0,?|y_1) = \frac{1}{2}A(u_1;y_0). 
	\]
	Next, define $f(u_0,y_0,?) = u_0$. Then, by~\eqref{eq:permutation} and~\eqref{eq:step7AP}, $A \circledast \ourP \permutation C_1$ as
	\begin{IEEEeqnarray*}{rCl}
		C_1(u_1 \oplus f(u_0,y_0,?) ; u_0,y_0, ?) &=& \frac{1}{2}A(u_1\oplus u_0; y_0) \\
							  &=& \frac{1}{2}A(u_0\oplus u_1; y_0) \\ 
							  &=& (A\circledast \ourP)(u_1;u_0,y_0,?). 
	\end{IEEEeqnarray*}

	To show that $A \inferior A \circledast \ourP $ we show that
	\[
	A \degraded C_2 \permutation A \circledast \ourP. 
\]
First, define $g(u_0,y_0,?) = u_0$. Then, by~\eqref{eq:permutation} and~\eqref{eq:step7AP},
	\begin{IEEEeqnarray*}{rCl}
		C_2(u_1;u_0,y_0,?) &=& (A \circledast \ourP)(u_1\oplus g(u_0,y_0,?);u_0,y_0,?) \\ &=& \frac{1}{2}A(u_1;y_0). 
	\end{IEEEeqnarray*}
	Let $Q(y_1|u_0,y_0,?) = 1$ if $y_1=y_0$ and $0$ otherwise.  Then, by~\eqref{eq:degraded}, indeed $A \degraded C_2$, as
	\begin{IEEEeqnarray*}{rCl} 
		\sol{\sum_{u_0,y_0}} C_2(u_1;u_0,y_0,?)Q(y_1|u_0,y_0,?)& =& \sum_{u_0} \frac{1}{2}A(u_1;y_1)\\  &=& A(u_1;y_1).
	\end{IEEEeqnarray*} 
	\emph{Step 8: $\ourS \circledast B \equiv \ourS$.} By \Cref{lem:AbetwennPandS}, $\ourS \circledast B \inferior \ourS$. It remains to show $\ourS \inferior \ourS \circledast B$. By~\eqref{eq:AcircledastBdef}, we have 
	\begin{IEEEeqnarray*}{rCl}
		(\ourS \circledast B)(u_1;u_0,?,y_1) &=& \ourS(u_0\oplus u_1 ; ?) B(u_1; y_1) \\ 
			   			     &=& \begin{cases} B(u_1; y_1), & u_1 = u_0, \\ 0, & \text{otherwise}.\end{cases}
	\end{IEEEeqnarray*}
	We now show that $\ourS \degraded C_3 \permutation \ourS \circledast B$. 
	Let $h(u_0,?,y_1) = u_0$. Then, by~\eqref{eq:permutation}, $C_3 \permutation \ourS \circledast B$ with 
	\begin{IEEEeqnarray*}{rCl} 
		C_3(u_1;u_0,?,y_1) &=& (\ourS \circledast B)(u_1\oplus h(u_0,?,y_1);u_0,?,y_1) \\ 
				   &=& (\ourS \circledast B)(u_1\oplus u_0;u_0,?,y_1) \\ 
				   &=& \begin{cases} B(u_1\oplus u_0; y_1), & u_1 \oplus u_0 = u_0, \\ 0, & \text{otherwise}\end{cases} \\ 
				   &=& \begin{cases} B(u_1\oplus u_0; y_1), & u_1  = 0, \\ 0, & \text{otherwise}\end{cases} \\ 
				   &=& \begin{cases} B(u_0; y_1), & u_1  = 0, \\ 0, & \text{otherwise}.\end{cases}  
	\end{IEEEeqnarray*}
	Next, by~\eqref{eq:degraded}, $\ourS \degraded C_3$ with $Q(?|u_0,?,y_1) = 1$.  

\emph{Step 9: $\ourP \circledast B \equiv B$.} By~\eqref{eq:AcircledastBdef}, we have $(\ourP \circledast B)(u_1; u_0, ?,y_1) = B(u_1;y_1)/2$.
We now show that $B \inferior \ourP \circledast B \inferior B$, which will prove the equivalence. 
	Indeed, $\ourP \circledast B \degraded B$, since in \eqref{eq:degraded} we can take
\[
	Q(u_0,?, y_1|y') = \begin{cases} \frac{1}{2}, & y' = y_1 ,\\
		0, & \text{otherwise},
	\end{cases}
\]
and $B \degraded \ourP \circledast B$, by \eqref{eq:degraded} with
\[
	Q(y'|u_0,?, y_1) = \begin{cases} 1, & y' = y_1 ,\\
		0, & \text{otherwise}.
	\end{cases}
\]

\emph{Step 10: $\ourS \circledast \ourS \equiv \ourS$, $\ourS \circledast \ourP \equiv \ourS$, $\ourP \circledast \ourS \equiv \ourS$, and $\ourP \circledast \ourP \equiv \ourP$.} These are special cases of Steps 6--9.
\end{IEEEproof}

\end{appendix}

\twobibs{
\bibliographystyle{IEEEtran} 
\bibliography{mybib.bib} 
}
{
\ifdefined\bibstar\else\newcommand{\bibstar}[1]{}\fi

}

\end{document}